\documentstyle[amsfonts, eqsecnum,cite,aps]{revtex}

\begin{document}
\title{Minimizing postulation in a senior undergraduate course in electromagnetism}
\author{Nelson P. Andion}
\address{Institute of Physics, Federal University of Bahia, \\
Salvador 40210-340, Brazil.\\
e-mail: andion@ufba.br.}
\maketitle

\begin{abstract}
An approach to the teaching of electromagnetism to senior undergraduate
students, designed for overcoming the fragmentation of the theory is
described. As usual it starts from the static case, but it is strictly based
on Helmholtz theorem of uniqueness of vector fields, on primary experimental
observations upon the sources of electric and magnetic fields, on Lenz's law
and on principles of superposition and reciprocity of interactions.
Thereafter, without further postulation, all laws and rules arise in a
procedure where electricity and magnetism parallel one another. Maxwell's
equations are built up on a step-by-step basis, showing that electromagnetic
field can be considered but a twofold linear vector field obeying the
simplest principles of physical theories. It is shown that every system of
units fits this formalism, provided suitable values are assigned to some
constants arising therein from superposition principle.
\end{abstract}

\section{Introduction\ \ }

Contrasting with earlier procedures\cite{AbrahamB} contemporary physical
sciences undergraduate teaching makes frequent use of some powerful
mathematical tools pertaining not only to calculus but also to the methods
of theoretical physics. Nevertheless, the axiomatic structure hardly takes
advantage of such powerful mathematical machinery. Indeed, in spite of their
extraordinary contribution in adapting the available mathematical basis to
pedagogical needs, most textbooks on electromagnetic theory intended to
senior undergraduate students approach the subject by means of the same
axioms used in books intended to freshmen: the laws of Coulomb, Amp\`ere,
Biot and Savart, Lorentz force equation, etc.\cite
{Slater,RMC,Hauser,HMarion,Lorrain}. On the other hand, outstanding effort
has been directed to minimizing the number of axioms of electromagnetic
theory, in the form of textbook chapters and papers. Unfortunately, they
require a mathematical background that is seldom mastered by students at
this course level, thus being labelled ``optional matter'' and often
omitted, although not disregarded\cite{Hauser2,Kobe}.

The pedagogical approach presented here stands as a mean term between the
conventional one and those highly synthetic works ultimately designed for
axiomatic purposes. It has been developed for a course in electromagnetic
theory to senior undergraduate students of physics, aiming to encompass the
mathematical apparatus and the theoretical concepts into an unitary
formalism which provided conciseness and axiomatic economy to the theory,
without loss of conceptual or technical knowledge: accordingly, the above
mentioned laws, which usually constitute axioms become, here, the consequence%
{\bf \ }of superposition and reciprocity principles applied to a vector
field arising from a scalar physical quantity, the electric charge. It
provides a better understanding of the theory and prepares the student for
the advent of Maxwell's equations, which are obtained on a step-by-step
basis without invoking the usual mosaic of postulates. Throughout the
formalism, electricity and magnetism parallel one another, showing their
affinity and common origin, even in the stationary case. After obtaining the
equation of electromagnetic induction Maxwell's equations are derived by
simply imposing the requirement of self-consistency to the previous set, a
procedure that leads to displacement current. Finally, the essentials of
time-varying field theory are discussed along with the derivation of the
wave equations on potentials and the Poynting theorem. A brief discussion on
systems of units for electromagnetic fields in a vacuum and its implications
in our general formalism ends its presentation.

The resulting formalism yields great emphasis on concepts, without
disregarding the need for training on calculation techniques envisaging the
solution of scientific problems. This procedure can better prepare the
student for dealing with physical concepts and develop his/her creativeness.

Our approach is founded on a threefold base: (i) a review of mathematical
concepts, referring to vector algebra and vector calculus, particularly the
theorem of Helmholtz; (ii) on the phenomenological realm, we begin with a
set of three elementary thought experiments: first, the {\it pendulum
electroscope}\cite{pelec}; second, the interaction between permanent
magnets; third, the interaction between an electric discharge and a compass,
to which the conservation of electric charge and Lenz's law are added; (iii)
from the conceptual realm, principles of superposition and reciprocity of
interactions. Following the general use of courses at this level we assume,
critically, the validity of Newtonian mechanics and Galilean
transformations. Besides, for reason of conciseness we present here only the
approach to electromagnetic systems in a vacuum (microscopic theory).
Macroscopic theory will be presented later\cite{Nelson}.

In Section II the mathematical basis of our approach is founded and further
combined, in Section III, with theoretical postulates and experimental
evidence for obtaining the field equations in the static case. Section IV
shows how a reciprocity can be invoked and justified in the formalism and
how Lorentz force is derived from it. Section V extends the field equations
to the time-dependent case, i.e., we arrive at Maxwell's equations and its
consequences. In Section VI a brief discussion is made about systems of
units and in Section VII we present our final remarks.

\section{Mathematical background}

In what mathematical support is concerned, we would like to point out the
statement of the theorem of Helmholtz\cite{Panofsky,Arfken}: {\it let }${\bf %
V}\left( {\bf r}\right) $ de{\it scribe a vector field. If we are given the
functions}${\bf \ }s{\bf (r)}${\it \ and }${\bf c(r)}${\it , such that,} 
\begin{equation}
\nabla \cdot {\bf V}=s{\bf (r)},  \label{eq1}
\end{equation}
{\it and } 
\begin{equation}
\nabla \times {\bf V}={\bf c(r)},  \label{eq2}
\end{equation}
{\it then }${\bf V}${\it \ is uniquely determined as:} 
\begin{equation}
{\bf V=-}\nabla \phi +\nabla \times {\bf A},  \label{eq3}
\end{equation}
{\it where }$\phi $ {\it and} ${\bf A}${\it \ are the scalar and vector
potentials, respectively, given by:} 
\begin{equation}
\phi ({\bf r})=(1/4\pi )\int \left[ s\left( {\bf r}^{\prime }\right) /\left| 
{\bf r-r}^{\prime }\right| \right] dV^{\prime }  \label{eq4}
\end{equation}
{\it and } 
\begin{equation}
{\bf A}({\bf r)}=(1/4\pi )\int \left[ {\bf c}\left( {\bf r}^{\prime }\right)
/\left| {\bf r-r}^{\prime }\right| \right] dV^{\prime }.  \label{eq5}
\end{equation}
According to Eqs. (\ref{eq3})-(\ref{eq5}), the resulting field has the form 
\begin{eqnarray}
{\bf V}\left( {\bf r}\right) &=&(1/4\pi )\int [s\left( {\bf r}^{\prime
}\right) \left( {\bf r-r}^{\prime }\right) /\left| {\bf r-r}^{\prime
}\right| ^3]dV^{\prime }  \label{eq2603.1} \\
&&\ +(1/4\pi )\int [{\bf c}\left( {\bf r}^{\prime }\right) \times \left( 
{\bf r-r}^{\prime }\right) /\left| {\bf r-r}^{\prime }\right| ^3]dV^{\prime
}. \nonumber
\end{eqnarray}
All integrals above are taken over the whole space. This theorem is a
powerful tool for developing the theory, because establishing $s$ and ${\bf c%
}$ in Eqs. (\ref{eq1}) and (\ref{eq2}), leads to the unique determination of 
${\bf V}$. Thus, the search for those two functions constitutes an
advantageuos starting point for developing a theory of the field ${\bf V.}$
Furthermore, Eqs. (\ref{eq4})-(\ref{eq2603.1}) show that, from the
mathematical standpoint, $s\left( {\bf r}^{\prime }\right) dV^{\prime }$ and 
${\bf c}\left( {\bf r}^{\prime }\right) dV^{\prime }$ play the role of field
sources located at point ${\bf r}^{\prime }$ and having scalar and vector
nature, respectively. When applied to the field vectors ${\bf E}$ and ${\bf B%
}$, this theorem provides not only a precise correlation between these
fields and their sources, but also the proper definition of other physical
quantities as, for instance, the magnetic vector potential. In addition, it
has a number of corollaries\cite{Panofsky2} which provide better
understanding of the theory clarifying, particularly, the origin of the
magnetic scalar potential\cite{Nelson2}. On the other hand, it has to be
pointed out that, for practical purposes, the theorem use is almost entirely
restricted to the stationary case, for in the general time-dependent
situation the finite character of the velocity of interactions impedes an 
{\it a priori }determination of the right-hand side term in Eqs. (\ref{eq1})
and (\ref{eq2}). In other words, despite being useful for theoretical
inspection of time-dependent fields, Helmholtz theorem cannot afford the
complete solution for them. Furthermore, in electromagnetism we deal with a
twofold vector field, a field described by vectors ${\bf E}$ and ${\bf B}$
which can be depicted by independent field equations in the static case but
not in the general (time-dependent) case, because of their common origin in
the dynamics of electric charges. Therefore our starting point must be the
field originated from stationary source configurations, i.e., the case when
time is ignorable and the non-instantaneous nature of interactions is
irrelevant. It is underlined that the solution for the time-dependent case -
as well as its analysis - will be worked out after establishing the general
equations, Maxwell's equations.

A second point to be stressed on the mathematical foundations of the
formalism is the uniqueness of products involving vectors, as long as only
linear operations are taken into account. Accordingly, it can be shown by
simple means that:

\begin{enumerate}
\item  The multiplication of a vector {\bf v }by a scalar c is the only
linear operation leading from the pair $\{c,{\bf v}\}$ to a vector quantity.

\item  The usual scalar multiplication of two vectors ${\bf u}$ and ${\bf v}$
is the only linear operation leading from them to a scalar quantity.

\item  Analogously, the vector multiplication of an ordered pair of vectors $%
\{{\bf u,v}\}$ is the only linear operation leading from this pair to a
vector quantity.\cite{n11}
\end{enumerate}

\section{Equations of the static field}

Assuming that electric and magnetic fields have vector nature, our formalism
starts from the determination of the analogues of Eqs. (\ref{eq1}) and (\ref
{eq2}) for both these fields. There are powerful theoretical arguments which
can be used for attaining that aim but, in order to weaken our set of
axioms, we prefer to approach the subject by means of some trivial
experimental concepts and facts. It wouldn't be important describing here a
detailed experiment. For our purpose a simple enumeration of the
experimental goals and the possibility of reaching them is enough. These
goals are proving:

\begin{enumerate}
\item  that the electric field is irrotational;

\item  that the magnetic field is divergenceless;

\item  the equivalence between permanent magnets and electric discharges as
sources of magnetic field.
\end{enumerate}

{\it Electricity: }Electric interactions are here {\it defined }by means of
well known elementary experiments consisting of rubbing some bodies against
wool, fur or certain fabrics and observing that they attract pieces of paper
and other light objects. Electrometers and pendulum electroscopes are among
the useful measuring instruments. At this stage of development we also
introduce the concept of insulators and conductors, as enlightened by the
concept of conduction, which can be easily checked from the experimental
standpoint. Then a Van de Graaff generator is presented, its operation being
described in terms of the preceding concepts and phenomena. Electric field,
represented by the vector ${\bf E}$, is defined as the responsible for the
electric interaction and its sources are shown to have scalar nature. This
fact can be demonstrated by charging a spherical metal ball - using a Van de
Graaff generator, for instance - and observing that a rotation around an
arbitrary axis passing across its center doesn't change its field, i.e., the
attraction/repulsion it causes on, say, the{\bf \ }pendulum electroscope.
Thus we prove that the electric field is irrotational, since we must have 
\begin{equation}
\nabla \times {\bf E}=0,  \label{eq13a}
\end{equation}
in order to cope with the non-vector nature of its sources. On the other
hand, taking into account that, in Eq. (\ref{eq1}), $s{\bf (r)}$ is the
scalar source density, we can write 
\begin{equation}
\nabla \cdot {\bf E=}k_1\rho \left( {\bf r}\right) ,  \label{eq14a}
\end{equation}
being $\rho \left( {\bf r}\right) $ the electric charge density and $k_1$ a
constant to be established according to a system of units.

{\it Magnetism:} In a similar manner, magnetism is defined as the
interaction between permanent magnets at rest. A compass may be used as
measuring instrument. The tendency to alignment of the compass in a
preferred direction is observed and assigned to an external field, due to
the Earth. It is neglected as compared to the magnetic field created by the
magnet. Alternatively, a {\it double magnetic needle}, as that used by
Faraday for neutralizing the Earth's magnetic field\cite{Purcell}, can be
used in place of the common compass. Magnetic field, represented by the
induction vector ${\bf B}$, is the responsible for magnetic interaction. Its
sources are proven to have vector nature, as their orientation affects the
field they produce, according to the orientation and reorientation of the
compass under such field. That these sources don't have a scalar component
can be observed by superposing two identical permanent magnets in opposite
directions and noticing that their fields vanish\cite{n13} by opposition of
the vector sources. If there were a scalar component of the field source it
would, instead, double its contribution to the field as the magnets were
superposed, regardless of the common axis of their directions. Therefore, we
must have: 
\begin{equation}
\nabla \cdot {\bf B}=0.  \label{eq15a}
\end{equation}

In the third and last part of the thought experiment, a Van de Graaff
generator supplies charge to a {\it rough capacitor}\cite{n14} which is then
conveniently discharged by means of a conducting wire. The resulting
magnetic field is detected by a compass. Thus we conclude that the motion of
electric charges is equivalent to permanent magnets, in the sense that it
also gives rise to a magnetic field. Although remarking that not all
magnetic fields are originated by moving electric charges, we claim, in view
of the referred experimental observation and our definition of magnetic
interaction that, from the mathematical standpoint, magnetic fields can {\it %
always} be seen as being so originated. In other words, it is always
possible to find distributions of electric currents leading to a prescribed
field, whatever its true sources\cite{n15}.

Before deriving the last equation on the vector fields, namely, the analogue
of Eq. (\ref{eq2}) for vector ${\bf B}$, we shall work out the hydrodynamic
picture of electric charge flows. We recall those concepts referring to mass
flows, as density, flux, etc., and establish, as usual, the charge density 
\begin{equation}
\rho =Nq  \label{eq16a}
\end{equation}
and the current density 
\begin{equation}
{\bf J}=Nq{\bf v},  \label{eq17a}
\end{equation}
for a system of moving charged particles, being $N$ the number of particles
per unit volume, $q$ the charge of each particle and ${\bf v}$ its drift
velocity. Conservation of electric charge is then imposed, leading to
continuity equation, 
\begin{equation}
\nabla {\bf \cdot J}+(\partial \rho /\partial t)=0.  \label{eq18a}
\end{equation}
An important remark is that the hydrodynamic picture is enough for
macroscopically describing the steady flow of charges. Therefore, $\rho {\bf %
\ }$and ${\bf J}$, given by Eqs. (\ref{eq16a}) and (\ref{eq17a}), are the
only quantities needed for describing this flow and, consequently, all its
effects. Whereas $\rho $ describes the charge configuration, ${\bf J}$
describes its motion; when combined they describe a sort of electrodynamic
state of the system, similarly to what occurs in strictly dynamic flows. As
long as the vorticity of ${\bf B}$ is considered as deriving from charge
flows, it {\it must} be a function of ${\bf J}$. Furthermore, for obeying
superposition principle it must be a {\it linear} function of ${\bf J}$.
Thus we have 
\begin{equation}
\nabla \times {\bf B}=k_2{\bf J(r),}  \label{eq19a}
\end{equation}
where, as for $k_1$, $k_2$ will be evaluated in accordance with the chosen
system of units. Equation (\ref{eq19a}) completes the aimed set of equations
on ${\bf E}$ and ${\bf B}$. Obtaining ${\bf E}$ and ${\bf B}$ is now
straightforward. According to Eqs. (\ref{eq2603.1}), (\ref{eq13a}), (\ref
{eq14a}), (\ref{eq15a}) and (\ref{eq19a}) we have: 
\begin{equation}
{\bf E}({\bf r)}=(k_1/4\pi )\int [\rho \left( {\bf r}^{\prime }\right)
\left( {\bf r-r}^{\prime }\right) /\left| {\bf r-r}^{\prime }\right|
^3]dV^{\prime }  \label{eq20a}
\end{equation}
and 
\begin{equation}
{\bf B}({\bf r)}=(k_2/4\pi )\int [{\bf J}\left( {\bf r}^{\prime }\right)
\times \left( {\bf r-r}^{\prime }\right) /\left| {\bf r-r}^{\prime }\right|
^3]dV^{\prime }.  \label{eq21a}
\end{equation}
Finally we must underline a special feature of the field equations, namely,
the asymmetry shown by Eq. (\ref{eq14a}) as compared to Eq. (\ref{eq15a})
and by Eq. (\ref{eq13a}) as compared to Eq. (\ref{eq19a}), which amounts to
the magnetic field not having scalar sources, i.e., for it arising from the
motion of the electric field sources, the electric charges. It is tempting
to admit the converse, i.e., the existence of magnetic poles directly
originating a magnetic field as well as - by its motion - an electric field
and, consequently, the possibility of reversing and eliminating that
asymmetry. In practice, although magnets - our original source of magnetic
field - are endowed with north and south ``poles'', attempts to divide them
into their ``polar'' parts are unsuccessful: the resulting parts are again
magnets, with north and south ``poles''. However, since long the existence
of magnetic poles has been assumed as theoretically possible\cite
{Dirac,Dirac2} and, more recently, it became an imposition to certain
theories. Besides, a great number of experimental attempts to detect
magnetic poles have been performed, leading in a few cases to positive events%
\cite{PDG}, being the most outstanding of these the one reported by Cabrera%
\cite{Cabrera}. In view of the only positive experiments being cosmic rays
or matter searches, of unpredictable repeatability, as well as in most cases
having divergent interpretations, the existence of magnetic poles is yet an
open theme in scientific investigation. Due to the scope of this approach we
won't consider it therein.

\section{Superposition and reciprocity leading to Lorentz force}

Comparing Eqs. (\ref{eq1}) and (\ref{eq2}) with Eqs. (\ref{eq13a}), (\ref
{eq14a}), (\ref{eq15a}) and (\ref{eq19a}) and taking into account our
discussion about Helmholtz theorem in Section II, we conclude that $\rho dV$%
\ and ${\bf J}dV$\ are elementary sources of ${\bf E}$\ and ${\bf B}$,
respectively. Nevertheless we must consider the different meanings of the
word {\it source}. We may call elementary source of a certain field a minute
part of the material system which originates such field, or we may so
entitle the physical quantity, carried by that material element, which
measures its intensity as field source. In other words, by {\it source }we
mean the physical medium or else its measurable attribute as physical agent.
The latter is the choice when we talk about, say, ``a point charge $q$'' and
give its position instead of referring to ``a particle bearing an amount $q$
of charge'' located somewhere. It is a shorthand inherited from mathematics,
i.e., we use the mathematical concept of field source (see Section II) as a
shorthand description of the material source. On the other hand, one may ask
how such a particle or an elementary material system would respond to an
external field. One shall expect that, as long as it is a source of a given
field, it is also sensitive to the presence of an external field of the same
kind and, even more, that the physical quantity which measures its {\it %
intensity }as field source will, reciprocally, measure its {\it sensitivity }%
to that external field. Particularly, this implies that a material medium
only suffers the action of fields of the kind it originates. The above
argumentation may be summarized as the following qualitative reciprocity
principle:{\it \ an elementary material system suffers the action of an
applied field in the same measure as it originates a field of that kind.} In
association with superposition principle it brings immediately on Lorentz
force law. Indeed, let us initially calculate the force exerted by an
external electric field on the elementary source of another. According to
superposition principle stated, e.g., as {\it ``summing up causes leads to
summing up their effects''}, the resulting force has to be linear in both
physical quantities which measure its immediate causes: the intensity of the
external electric field, given by the field vector ${\bf E}_{ext}$, and the
elementary source sensitivity given by $\rho dV$, according to reciprocity
principle. In view of the discussion in Section II about products involving
vectors, the only way of combining those two quantities for obtaining a
linear result is the multiplication of the vector ${\bf E}_{ext}$ by the
scalar $\rho dV$. Thus we must have, for that force:

\begin{equation}
d{\bf F}_e=\alpha \left( \rho dV\right) {\bf E}_{ext},  \label{eq22a}
\end{equation}
being $\alpha $ another constant referring to the system of units.
Analogously, being the vector product the only possibility of linearly
combining two vectors (see, again, Section II), the magnetic force resulting
from the action of an external magnetic field, ${\bf B}_{ext}$, on an
elementary magnetic field source, ${\bf J}dV$, must be given by 
\begin{equation}
d{\bf F}_m=\beta \left( {\bf J}dV\right) \times {\bf B}_{ext},  \label{eq23a}
\end{equation}
being $\beta $ the last constant attached to a system of units. Now, in view
of Eq. (\ref{eq16a}), Eq. (\ref{eq22a}) takes the form 
\begin{equation}
d{\bf F}_e=\alpha \left( dn\right) q{\bf E}_{ext},  \label{eq25a}
\end{equation}
being 
\begin{equation}
dn=NdV  \label{eq26a}
\end{equation}
the number of particles in volume $dV$. It is immediately noticed from Eq. (%
\ref{eq25a}) that the force exerted by an electric field ${\bf E}$ on a
point charge $q$ is 
\begin{equation}
{\bf F}_e=\alpha q{\bf E}.  \label{eq27a}
\end{equation}
Analogously, introducing the expression of ${\bf J}$, given by Eq. (\ref
{eq17a}), into Eq. (\ref{eq23a}) we get: 
\begin{equation}
d{\bf F}_m=\beta \left( dn\right) q{\bf v}\times {\bf B}_{ext},
\label{eq29a}
\end{equation}
being $dn$ as defined by Eq. (\ref{eq26a}). Then, in the discrete limit,
when a charge $q$ moves with velocity ${\bf v}$ under a field ${\bf B}$, we
have 
\begin{equation}
{\bf F}_m=\beta q{\bf v}\times {\bf B}.  \label{eq30a}
\end{equation}
Summing up Eqs. (\ref{eq27a}) and (\ref{eq30a}) we get Lorentz force
equation: 
\begin{equation}
{\bf F}_{em}=q\left( \alpha {\bf E+}\beta {\bf v}\times {\bf B}\right) .
\label{eq31a}
\end{equation}
The above equation has been derived under the assumption that the particle
which suffers the force ${\bf F}_{em}$ belongs to a steady flow.
Nevertheless it must be remarked that such condition doesn't restrict the
type of motion of an {\it individual }particle in the flow. Indeed, it can
be shown that the particle may have any instantaneous acceleration - not to
mention its velocity - even in this case. Therefore, we arrive to the
conclusion that Eq. (\ref{eq31a}) is valid whatever the particle motion.

Introducing $\rho ^{\prime }=q^{\prime }\delta \left( {\bf r}^{\prime
}\right) $ into Eq. (\ref{eq20a}) and applying the result in Eq. (\ref{eq27a}%
) we obtain Coulomb's force between charges $q$ and $q^{\prime }:$%
\begin{equation}
{\bf F}_e=(\alpha k_1/4\pi )(qq^{\prime }{\bf r/}r^3),  \label{eq1704}
\end{equation}
showing that Coulomb's law is but a measurable result of the basic
conditions satisfied by electrostatic interaction: (i) superposition
principle, (ii) reciprocity principle and (iii) scalar nature of the
electrostatic field source. Particularly, the conservative nature of
electrostatic field is due to the scalar nature of the field source, as can
be noted in the derivation of Eq. (\ref{eq13a}). Analogously, for obtaining
the laws of magnetism it is only required to add to those principles the
assumption that the static magnetic field arises from the steady flow of
electric field sources. We may also obtain the interaction force of a pair
of parallel wires carrying currents $I$ and $I^{\prime }$ by combining, for
instance, Eq. (\ref{eq19a}) in integral form (Amp\`ere's circuital law) with
Eq. (\ref{eq23a}), and using the correspondence between a volume
distribution of charges and a current circuit: $\int *{\bf J}%
dV\leftrightarrow I\oint *d{\bf r}$ - where the asterisk stands for an
arbitrary factor (an scalar or a vector) and the multiplication sign, if
needed. There results, for the magnetic force per unit length of wire: 
\begin{equation}
{\bf F}_{m1}=-(\beta k_2II^{\prime }/2\pi d){\bf t,}  \label{magforce}
\end{equation}
where ${\bf t}$ represents the unit vector in a direction on the plane
containing the wires which is perpendicular to them and points outward.

The set of fundamental equations of static electric and magnetic
interactions is now accomplished. The choice of constants $k_1,k_2,\alpha $
and $\beta $ in Eqs. (\ref{eq14a}), (\ref{eq19a}) and (\ref{eq31a}) implies
defining a system of units. For performing applications and solving problems
at this stage of development one may ascribe these constants the values
pertaining to the preferred system of units as well as, correctly, the
positive sign to all them, and postpone the detailed analysis of different
possibilities of signs and values until obtaining Maxwell's equations and
analyzing its implications. Then it will be possible to restrict their signs
to be positive, find the only relation among them: $\alpha k_1/\beta
k_2=c^2, $ and determine the rules governing the systems of units used in
electromagnetism, as we do in Sections V and VI.

All laws and useful relations valid for static electric and magnetic
interactions follow from the equations formerly derived. Accordingly, Eqs. (%
\ref{eq14a}) and (\ref{eq19a}) are, respectively, Gauss' law and Amp\`ere's
circuital{\bf \ }law in differential form. As previously pointed out, the
Coulombian field and the laws governing magnetic phenomena follow from Eqs. (%
\ref{eq21a}) and (\ref{eq30a}). The meaning of electric potential and the
proof that it obeys Poisson's or Laplace's equation, the derivation of
relations which establish the electric and magnetic energy and energy
density, etc., also follow straightforwardly from the previous relations. In
brief, no concept or relation is omitted in our approach, although they may
appear in a different order.$_{}$

\section{Time-dependent fields}

In this Section we extend the analysis of electromagnetic phenomenology to
the time-dependent case, starting from electromagnetic induction. It can be
rigorously shown that it isn't an independent phenomenon. On the contrary,
the law of electromagnetic induction - Faraday's law - and, in general,
Maxwell's equations can be derived from the phenomenology of static fields
by suitably applying a Lorentz transformation to the corresponding quantities%
\cite{Hauser2,Kobe}. However, as we already noticed (Section I), such
derivation requires the knowledge of concepts and techniques seldom mastered
by students at this course level, but we can avoid further postulation by
adopting an approach\cite{Tamm} based on a Galilean transformation imposed
to the Lorentz force that a purely magnetic field exerts on charges at rest
on a circuit attached to a moving frame, as well as on the invariance of
physical laws, particularly Lorentz force equation. It is important to note
that, although Lorentz transformation would be the only rigorous choice, the
Galilean one leads to the same result, namely: 
\begin{equation}
\nabla \times {\bf E}=-(\beta /\alpha )(\partial {\bf B}/\partial t){\bf ,}
\label{eq1304.1}
\end{equation}
as the general form of the electromagnetic induction law, Faraday's law.
Constants $\alpha $ and $\beta $ arise here in accordance to our general
formalism.

Imposition of self-consistency to the set of Eqs. (\ref{eq14a}), (\ref{eq15a}%
), (\ref{eq19a}), (\ref{eq1304.1}) and the equation of continuity, Eq. (\ref
{eq18a}), leads to displacement current and to substitution of Amp\`ere's
law, Eq. (\ref{eq19a}), by 
\begin{equation}
\nabla \times {\bf B}=k_2{\bf J}+(k_2/k_1)(\partial {\bf E}/\partial t).
\label{eq1304.2}
\end{equation}
The unchanged equations plus the latter constitute the set of microscopic%
\cite{n22} Maxwell's equations.

For completing the picture given by this formalism we will analyze a few
outstanding consequences of Maxwell's equations: the wave equations on
electromagnetic potentials and the energy balance described by Poynting's
theorem.

Adopting a procedure analogous to the usual one\cite{Jackson} we derive,
from Eqs. (\ref{eq15a}) and (\ref{eq1304.1}), the electromagnetic vector and
scalar potentials, ${\bf A}_{em}({\bf r})$ and $\phi _{em}({\bf r})$,
respectively, related to the time-dependent field vectors according to: 
\begin{equation}
{\bf B(r)=}\nabla \times {\bf A}_{em}  \label{vectpot}
\end{equation}
and 
\begin{equation}
{\bf E}({\bf r)}=-\nabla \phi _{em}-(\beta /\alpha )(\partial {\bf A}%
_{em}/\partial t).  \label{scalpot}
\end{equation}
Applying to them the remaining Maxwell's equations, Eqs. (\ref{eq14a}) and (%
\ref{eq1304.2}), and imposing to the result the condition: 
\begin{equation}
\nabla \cdot {\bf A}_{em}+(k_2/k_1)(\partial \phi _{em}/\partial t)=0,
\label{lorentz}
\end{equation}
to be identified as the Lorenz condition in its general form, we are lead to
the wave equations 
\begin{equation}
\nabla ^2\phi _{em}-(\beta k_2/\alpha k_1)(\partial ^2\phi _{em}/\partial
t^2)=-k_1\rho  \label{scalwv}
\end{equation}
and 
\begin{equation}
\nabla ^2{\bf A}_{em}-(\beta k_2/\alpha k_1)(\partial ^2{\bf A}%
_{em}/\partial t^2)=-k_2{\bf J.}  \label{vectwv}
\end{equation}
In the absence of charges and currents the above equations reduce to
d'Alembert's equations in $\phi _{em}$ and ${\bf A}_{em}$, provided that $%
\beta k_2/\alpha k_1>0$. Otherwise those equations lead to decaying or
building-up solutions, without physical meaning. Moreover, in the meaningful
case $\alpha k_1/\beta k_2$ represents the square of the electromagnetic
wave velocity, which is experimentally known to equal the velocity of light, 
$c$, in view of light waves belonging to the electromagnetic spectrum. Thus
we have, 
\begin{equation}
\alpha k_1/\beta k_2=c^2>0.  \label{wavel}
\end{equation}
The physically meaningful solutions of Eqs. (\ref{scalwv}) and (\ref{vectwv}%
) are the well-known retarded potentials.

On the other hand, applying to Eqs. (\ref{eq1304.1}) and (\ref{eq1304.2}) a
treatment analogous to the one adopted by Landau\cite{Landau} we are lead to
the energy relation: 
\begin{equation}
\partial \left\{ \int_V(1/2)\left[ (\alpha /k_1)E^2+(\beta /k_2)B^2\right]
dV+\sum K\right\} /\partial t=-\oint_S{\bf S\cdot n}dS,  \label{poynthm}
\end{equation}
being $V$ the volume enclosed by surface $S.$ The summation $\sum K$
represent the total kinetic energy of the enclosed charged particles or,
alternatively, their total energy, including their rest masses (these are
constants, thus their inclusion don't alter the final value of the
derivative). The vector ${\bf S}$ in the right-hand side is Poynting vector,
defined here as{\bf :} 
\begin{equation}
{\bf S=}(\alpha /k_2)({\bf E\times B{\Bbb )}.}  \label{poyntvec}
\end{equation}
If $V$ is extended to comprise the whole space and the charges and currents
form a bounded system, Eq. (\ref{poynthm}) reduces to 
\begin{equation}
d\left\{ \int (1/2)\left[ (\alpha /k_1)E^2+(\beta /k_2)B^2\right] dV+\sum
K\right\} /dt=0.  \label{enblnce}
\end{equation}
This equation tells us that the quantity into the outer brackets is
conserved. Being $\sum K$ the kinetic (or the total) energy of the particles
in the system, we conclude that the quantity into outer brackets represents
the mechanical energy of the system of particles, being the integral - now
taken over the whole space - the energy stored in the electromagnetic field.
Accordingly, the integrand 
\begin{equation}
\omega =(1/2)\left[ (\alpha /k_1)E^2+(\beta /k_2)B^2\right]  \label{endens}
\end{equation}
is associated with the energy density of the field. Thus, Eq. (\ref{poynthm}%
) represents the energy balance of the system - Poynting's theorem - being $%
{\bf S}$ a measure of the energy flow per unit area and per unit time. Yet
all this familiar digression has been performed up to the determination of
individual signs on the constants of the set $\left\{ k_1,k_2,\alpha ,\beta
\right\} $, which we are now in a position to accomplish{\bf , }thus
confirming the positive value we ascribed them {\it a priori} (see Section
IV).{\it \ }

From Eq. (\ref{wavel}) we conclude that the products $\alpha k_1$ and $\beta
k_2$ have the same sign. Consequently, the same holds for $\alpha /k_1$ and $%
\beta /k_2$. Thus the field energy, given by the integral in Eq. (\ref
{enblnce}), will have this common sign. In other words, it will be positive
or negative according to these products (ratios) having the positive or the
negative sign. However the field energy can't be negative. In fact, in
deriving Eq. (\ref{enblnce}) no relation has been assumed between the charge
of particle-sources and their masses or velocities. Therefore, under the
hypothesis of negative field energy we could imagine a system of particles
with sufficiently great values of charges and sufficiently small values of
masses and velocities, so that its total relativistic energy, i.e., its rest
mass, given by the quantity into the outer brackets in Eq. (\ref{enblnce}),
would be negative, which is obviously forbidden\cite{relativity}. Thus we
must rule out negative signs from the products (ratios) $\alpha k_1$ $%
(\alpha /k_1)$ and $\beta k_2$ $(\beta /k_2)$. It results that the ordered
set $\left\{ k_1,k_2,\alpha ,\beta \right\} $ can only bear signs according
to the schemes: (1) $\left\{ ++++\right\} $; (2) $\left\{ +-+-\right\} $;
(3) $\left\{ -+-+\right\} $, or (4) $\left\{ ----\right\} $, each sign
belonging to the constant which occupies the same site into the brackets.
Now we proceed to show that among the four cases, those which differ by an
overall sign inversion - case 1 as compared to case 4 and case 2 as compared
to case 3 - are equivalent. Indeed, Maxwell's equations in absence of
charges and currents, i.e., for $\rho =0$ and ${\bf J}=0$, don't change by
an overall inversion of the signs on these constants (cf. Eqs. (\ref{eq14a}%
), (\ref{eq15a}), (\ref{eq1304.1}) and (\ref{eq1304.2})). Neither are
Coulomb's law, Eq. (\ref{eq1704}), and the force between current circuits
(see, e.g., Eq. (\ref{magforce})) changed under such an operation. This
means that both the {\it field-field }relation and the {\it %
particle-particle }interaction remain unchanged under a general inversion of
signs on those constants. The only change is noted in the {\it %
particle-field }relation, but it isn't significant here. It represents an
inversion on charge sign: we would name positive what is normally considered
negative and vice-versa. In view of this analysis the number of cases are
reduced, even more, to a half. Accordingly, we choose $k_1>0$ - which
amounts to establishing that a positive source gives rise to divergent lines
of force and conversely - so eliminating cases 4 and 3 in view of cases 1
and 2, respectively. Now, from Eq. (\ref{eq1304.1}) we notice that only case
1 obeys {\it Lenz's law}; case 2, for which $\alpha $ and $\beta $ have
different signs, doesn't follow that law and must also be discarded. Thus we
conclude that, as long as electromagnetism is described by our postulates
all constants of the set $\left\{ k_1,k_2,\alpha ,\beta \right\} $ are
positive\cite{magpoles}. We have thus confirmed the signs we anticipated for
those constants in Section IV, as well as the relation given by Eq. (\ref
{wavel}), among them.

As for the results we obtained in the static case, the above results
represent the core of electromagnetic theory for time-dependent fields.
Indeed, Maxwell's equations and Lorentz force equation are enough for
describing all phenomena related to electromagnetic waves, electrodynamics
of moving particles, etc. As a whole, the previous analyses of static and
time-dependent cases form an axis from which many specific subjects and
problems branch, whose formulation entails a corresponding number of
interesting approaches and solutions yet not fitting into the space here
available. However, in view of its importance and general applicability,
that of systems of units represents a subject worth while analyzing here.
Particularly, when suitably employed the results of such analysis unveil the
generality of this formalism, under which problems and applications can be
worked out regardless of the system of units used in their formulation.

\section{Systems of units}

Theoretical and historical reasons converged to transform the systems of
units in electromagnetism into one of the most entangled theme of the
undergraduate physics course. Theoretical freedom to choose, represented by
the four constants previously defined - having only one required relation
linking them - enhanced by other undetermined constants related to fields in
material media, was the background for early definition of {\it practical
units }and their dictating the establishment and renewal of several systems,
leading eventually, in 1960, to the SI ({\it Syst\`eme International
d'Unit\'es})\cite{Jordan-Balmain}. Although ending in a scientific system
this process betrays its origin through the complex interrelations among the
resulting units and between them and those of systems created directly for
scientific purposes - particularly the {\it gaussian system}, the most used
among the latter. Determination of Maxwell's equations governing fields in
material media requires an extended discussion about the properties of those
materials as well as about coherence between the corresponding electric and
magnetic constitutive equations. We shall avoid it here and only discuss the
subject of system of units envisaging applications to fields in a vacuum.

Units of electromagnetic quantities are linked to the mechanical ones by Eq.
(\ref{wavel}), or by Eqs. (\ref{eq1704}) and (\ref{magforce}), or by other
equivalent relations. In other words, those units aren't independent from
mechanical units but, instead, shall be defined in connection to them. Thus,
systems of electromagnetic units are based on mechanical systems of units,
CGS and MKS, and accomplish them. In conclusion, they must obey two
requirements:

\begin{enumerate}
\item  being attached to the CGS or the MKS system of units;

\item  the corresponding constants being interrelated according to Eq. (\ref
{wavel}).
\end{enumerate}

As we pointed out, there are two commonly used systems: the SI, linked to
the MKS system and having constants given by $\left\{ 1/\varepsilon _0,\mu
_0,1,1\right\} $ - in the order previously defined - and the gaussian
system, linked to the CGS and having the constants $\left\{ 4\pi ,4\pi
/c,1,1/c\right\} $. Note that both obey Eq. (\ref{wavel})\cite{SI}. Whenever
needed, our equations can be readily adapted to one system or another by
substituting the values of these constants in them. Nevertheless we believe
that this shall be done only in applications, leaving the constants and the
whole theory in their general form insofar as theoretical developments are
aimed.

\section{Final remarks}

Starting from the theorem of Helmholtz, which establishes the conditions for
uniqueness of vector fields, and supported by elementary experimental
concepts on the field source, a theoretical basis is readily accomplished as
differential equations for the stationary state field vectors, ${\bf E}$ and 
${\bf B}${\bf , }representing the basic laws regarding electrostatic and
magnetostatic fields. Then, Lorentz force arises from the principles of
superposition and reciprocity of electromagnetic interactions through a very
simple reasoning, and the law of electromagnetic induction - Faraday's law -
is also obtained from them by imposing a Galilean transformation of
reference frame on forces due to stationary magnetic fields. As a
consequence of self-consistency being imposed to the previous set of
equations, Maxwell's equations are obtained up to the determination of four
constants whose signs are uniquely defined with the help of Lenz's law and
whose values are linked by a single relation, derived from the experimental
fact of light waves belonging to the electromagnetic spectrum. In short,
this approach has the following essential features:

\begin{enumerate}
\item  The field equations are proposed, from the beginning, in Maxwell's
format, i.e., in terms of divergence and curl differential operators, being
continuously adjusted until transforming into Maxwell's equations.

\item  Beginning with the theorem of Helmholtz, it clarifies some concepts
as, for instance, those of field source, both in its mathematical and
physical meaning, and vector magnetic potential.

\item  The whole theory follows from trivial principles and from elementary
experimental observations.

\item  Along its development, electric and magnetic phenomenology are always
presented side by side, and their affinity is clearly shown even for static
fields. The presentation thus requires less expenditure of time in class.

\item  Uniqueness of Maxwell's equations and their consequences are
unambiguously derived from those primary principles.

\item  Alternatively, the formalism would apply to other kinds of vector
fields obeying different primary principles, restricting their possible
number and properties, according to the signs and values of constants $%
\left\{ k_1,k_2,\alpha ,\beta \right\} $.

\item  Every system of units fits the formalism, provided that suitable
values are assigned to those constants.
\end{enumerate}

In essence, the formalism was based on primary experimental concepts,
through which we have shown that an scalar quantity - the {\it electric
charge - }gives rise to both the electric and the magnetic fields, being the
former a consequence of its presence and the latter one of its motion. Thus,
electromagnetic laws and rules represent the simplest picture Nature can
stand for those vector field\cite{n23}. However we must remark that it was
the available mathematical apparatus which provided these conclusions, thus
it is an underutilization of technical resources not reaching them in a
senior undergraduate course.

The emphasis on calculation techniques - brought to undergraduate teaching
in physics some forty years ago - seems to have looked at the urging demand
for development of new technologies, but it can do much more. It provides
the means for presenting to students a sequential theory of
electromagnetism, with closely intertwined concepts, thus enlightening its
overall picture and overcoming its usual fragmentation. In other words, the
final result may be a clear and unitary course.

\section*{Acknowledgments}

I am greatly indebted to Olival Freire Jr. and Saulo Carneiro for their
careful reading of a previous version of the manuscript and their invaluable
suggestions, and to Ademir Santana and Marcelo Moret for their encouraging
help on solving specific problems. I also address my thanks to the students
whose comments and even doubts were very worthy in the process of
transforming the initial ideas into a defined pedagogical formalism.

\newpage\


\begin{references}
\bibitem{AbrahamB}  M. Abraham and R. Becker, {\it The Classical Theory of
Electricity and Magnetism} (Blackie \& Son, London, 1950).

\bibitem{Slater}  J. C. Slater and N. H. Frank, {\it Electromagnetism} (
McGraw-Hill, New York, 1947 ).

\bibitem{RMC}  J. R. Reitz, F. J. Milford, and R. W. Christy, {\it %
Foundations of Electromagnetic Theory} (Addison-Wesley, Reading, 1980).

\bibitem{Hauser}  W. Hauser, {\it Introduction to the Principles of
Electromagnetism }(Addison-Wesley, Reading, 1971).

\bibitem{HMarion}  M. A. Heald and J. B. Marion, {\it Classical
Electromagnetic Radiation} (Saunders College Publishing, Orlando, 1995).

\bibitem{Lorrain}  P. Lorrain and D. R. Corson, {\it Electromagnetic Fields
and Waves} (Freeman, San Francisco, 1970).

\bibitem{Hauser2}  See Ref. 4, pp. 232-243.

\bibitem{Kobe}  D. H. Kobe, ``Generalization of Coulomb's law to Maxwell's
equations using special relativity,'' Am. J. Phys. {\bf 54} (7), 631-636
(1986).

\bibitem{pelec}  {\it Pendulum electroscope}: a device consisting of a tiny
ball whose core is made out of a light material as, for example, cork,
covered by a thin conducting layer, the whole being suspended from a support
by a fine thread of silk or another light insulating fiber. Under influence
of the nonuniform electric field of a bounded system of charges, the ball
suffers electric induction and moves toward that source. This movement can
be reversed by neutralizing the opposite induced charge through contact with
the inducing source.

\bibitem{Nelson}  N. P. Andion, {\it Minimizing postulation in a senior
undergraduate course in electromagnetism. Part II: fields in material media}
(to be published).

\bibitem{Panofsky}  W. Panofsky and M. Phillips, {\it Classical Electricity
and Magnetism} (Addison-Wesley, Reading, 1955), pp. 1-5.

\bibitem{Arfken}  G. Arfken, {\it Mathematical Methods for Physicists}
(Academic, New York, 1970), 2nd. ed., pp. 66-70.

\bibitem{Panofsky2}  See Ref. 11, pp. 5 and 6.

\bibitem{Nelson2}  See Ref. 10.

\bibitem{n11}  Among ``vector quantities'' we also count the pseudo-vectors.

\bibitem{Purcell}  E. M. Purcell, {\it Electricity and Magnetism}
(McGraw-Hill, New York, 1963), 2nd.ed., pp. 226-228.

\bibitem{n13}  Unless for a short-range residual field due to quadrupoles
and higher-order multipoles, which could at this moment be justified by the
non-interpenetration of the magnets.

\bibitem{n14}  We mean a kind of capacitor whose functioning can be easily
explained and understood - in terms of the concepts previously established
here - as for example a Leyden bottle. Thus we avoid introducing a
``black-box'' into the experiment.

\bibitem{n15}  At this point we must refer to fields originating from {\it %
intrinsic }magnetic moments of microscopic origin. The assumed equivalence
between magnets and currents is confirmed later when, as usual, magnets are
shown to behave as volume and surface distributions of electric current
(magnetization currents).

\bibitem{Dirac}  P. A. M. Dirac, {\it Quantised Singularities in the
Electromagnetic Field}, Proc. Roy. Soc. London {\bf A133}, 60-72 (1931).

\bibitem{Dirac2}  P. A. M. Dirac, {\it The Theory of Magnetic Poles}, Phys.
Rev.{\bf 74}, 817-830 (1948).

\bibitem{PDG}  Particle Data Group, {\it Review of Particle Physics}, Phys.
Rev. D {\bf 54}, 685-687 (1996).

\bibitem{Cabrera}  B. Cabrera, {\it First Results from a Superconductive
Detector for Moving Magnetic Monopoles}, Phys. Rev. Lett. {\bf 48},
1378-1381 (1982).

\bibitem{Tamm}  I. E. Tamm, {\it Foundations of the Theory of Electricity }%
[Spanish version] (Mir, Moscow, 1979), pp. 381-389.

\bibitem{n22}  Deriving the general (macroscopic) equations requires a
consistent development of field equations in material media, which would
extend too much this paper (see Ref. 10){\bf .}

\bibitem{Jackson}  J. D. Jackson, {\it Classical Electrodynamics}, 2nd. ed.
(John Wiley and Sons, New York, 1975) pp. 219-220.

\bibitem{Landau}  L. Landau and E. Lifshitz, {\it The Classical Theory of
Fields} (Addison-Wesley, Cambridge, Mass., 1951) pp. 78-80.

\bibitem{relativity}  Although being Galilean in general, our approach can
easily bear such a relativistic digression, based on a matter of fundamental
scientific knowledge.

\bibitem{magpoles}  On the other hand it can be shown, by interchanging the
roles of electric and magnetic fields in the theory and reinterpreting
Lenz's law, that the theory of an electromagnetic field arising analogously
from {\it magnetic poles }would obey case 2. Thus, the generalized theory
comprises both physically meaningful cases!

\bibitem{Jordan-Balmain}  E. C. Jordan and K. G. Balmain, {\it %
Electromagnetic Waves and Radiating Systems}, 2nd.ed. (Prentice-Hall, Inc.,
New Jersey, 1968), pp. 22-23.

\bibitem{SI}  We recall that $\varepsilon _0\mu _0=1/c^2.$

\bibitem{n23}  Its apparent intricacy is due to a sort of logical inversion:
the usual choice of following historical chronology when developing a course
in electromagnetic theory leads to reversing the roles of suitable premises
and suitable consequences. So are, for example, the statements of fields and
potentials in the conventional formalism.
\end{references}
\end{document}